\documentclass[prl,twocolumn]{revtex4}
\usepackage{graphics}
\usepackage{graphicx}
\usepackage{amsmath}
\usepackage{amstext}
\usepackage{xcolor}
\usepackage{}
\usepackage{amssymb}
\usepackage{hyperref}
\newcommand{\la}{\label}

\newcommand{\e}{{\epsilon_{kij}}}

\newcommand{\bfj}{{\bf j}}
\newcommand{\rhot}{ \tilde{\rho}}
\newcommand{\lt}{ \tilde{\lambda}}
\newcommand{\cho}{ \tilde{\chi}_1}
\newcommand{\cht}{ \tilde{\chi}_2}

\newcommand{\bfn}{{\vec{\bf  n}}}
\newcommand{\bfe}{{\vec{\bf  u}}}

\newcommand{\bfC}{{\bf C}}
\newcommand{\bfCt}{\tilde{\bf C}}

\newcommand{\f}{\frac}
\newcommand{\ndn}{\bfn \cdot \partial_i \bfn \times \partial_j \bfn}
\newcommand{\ede}{\bfe \cdot \partial_i \bfe \times \partial_j \bfe}

\newcommand{\bfa}{{\bf A}}

\newcommand{\be}{\begin{equation}}
\newcommand{\ee}{\end{equation}}
\newcommand{\ba}{\begin{eqnarray}}
\newcommand{\ea}{\end{eqnarray}}
\newcommand{\bastar}{\begin{eqnarray*}}
\newcommand{\eastar}{\end{eqnarray*}}

\begin{document}
\title{Non-Meissner electrodynamics and knotted solitons  \\ in  two-component superconductors}
\author{Egor Babaev}
\affiliation{
Physics Department, University of Massachusetts, Amherst MA 01003, USA\\
Department of Theoretical Physics, The Royal Institute of Technology 10691 Stockholm, Sweden}

\date{\today }
\begin{abstract}
I consider electrodynamics and the problem of knotted solitons in two-component superconductors.
Possible existence of knotted
solitons in multicomponent superconductors was predicted several years ago. However
their basic properties and stability in these systems remains an outstandingly difficult 
question both for analytical and numerical treatment.
 Here I  propose
a new  perturbative approach to treat self-consistently all the degrees of freedom in the problem. I  show that there exists a length scale for a
Hopfion texture where the electrodynamics of a two-component superconductor is dominated by a self-induced Faddeev
term, which is a stark contrast to the Meissner electrodynamics of single-component systems.
I also show that at certain short length scales knotted solitons in two-component Ginzburg-Landau model are not described by a Faddeev-Skyrme-type model and are unstable. However these solitons
can be stable at some intermediate length scales. I argue that   configurations with a high topological charge may be 
more stable in this system
than low-topological-charge configurations.
In the second part of the paper I  discuss qualitatively different physics of the stability of knotted solitons in
a more general Ginzburg-Landau model and point out the physically relevant terms which enhance or suppress stability
of the knotted solitons. With this argument it is demonstrated that the generalized Ginburg-Landau model 
possesses stable knotted solitons.
\end{abstract}
\maketitle

Quintessential and unusual properties of a quantum fluid consisting of a large number 
of particles
can be described at macroscopic length scales by a simple complex scalar field 
$\Psi ({\bf r})$, or in the case of a multicomponent quantum fluid by a multiplet
 of  complex fields  $\Psi_i({\bf r}), \  i=1,2,..,N$.
The fact that this simple description 
 is possible implies that the flow of the macroscopically
large number of particles comprising a quantum fluid is severely constrained.
The constraint in question is the superflow quantization 
condition which originates from the single-valuedness of the complex fields $\Psi_i ({\bf r})$. 
 In a physical situation where there is a superflow locally in space
 this flow should satisfy the single-valuedness condition of the condensate 
 wave function. Thus it can be created by exciting  vortex loops with quantized
  superfluid velocity circulation, or magnetic flux.
  Consequently many key properties of  quantum fluids depend dramatically on properties 
of the vortex loops.
In presently known and well investigated superfluids and superconductors the main characteristic
which vortex loops have in common is the fact that their energies depend monotonically on the 
loop sizes.
Indeed this fact is very important for physics of how a quantum fluid restores symmetry 
(via entropy driven proliferation of vortex loops), reacts to a quench (via a relaxation 
of a quench induced vortex loops), for physics of superfluid turbulence,  etc.
 All these properties would be quite dramatically altered if some quantum fluid  
 would allow vortex loops with  nonmonotonic energy dependence on the loop  size 
 (i.e. if the energy of the vortex loop would grow
 not only if the vortex loop expands but also if it shrinks below some characteristic size). 
  In these cases quench-induced  defects would be protected from decay by a potential barrier resulting in large-scale 
remnant post-quench vorticity. It may also produce hysteretic behavior in entropy-generated
 topological defects and thus 
change the order of the superconducting phase transitions etc.
Essentially in many respects this would lead to a new type of superfluid behavior, but the outstanding question is whether any quantum fluids can, in principle, support such defects.

The  model which is considered below applies, with some modifications, to  a large variety
multicomponent systems such as multicomponent electronic condensates (e.g. two-band superconductors \cite{frac}),
physics of the projected multicomponent quantum  metallic fluid of hydrogen or its isotopes under high compression \cite{Nature},
similar situation may arise in spin-triplet superconductors \cite{tripl}, similar models were also discussed in
the context of neutron star interior \cite{ns,Jones}.
Besides that in condensed matter physics there is a growing interest in systems where $SU(2)$ Ginzburg-Landau 
functional appears as an effective model \cite{DQC,DQC1}. In that context it is indeed important to understand the basic properties of topological defects in $SU(2)$ superconductor
in order to understand fluctuations and critical behavior in these systems. 

In \cite{PRL02} and also in \cite{tripl,ns} it was conjectured that some multicomponent superconductors may support  defects in the form of loops or knots
which energy would grow if such a vortex shrinks. However 
this energy scaling and the question of stability of these defects in these systems turned out to
 be extremely difficult, and in spite of multiple attempts to solve the problem no conclusive 
 results were found so far.
In this work I use a new approach to show that indeed in certain cases,  condensed matter systems 
should allow topological defects endowed with such properties. 

The question of the existence of topological solitons in the form of loops or knotted loops
which  energy is a nonmonotonic function of size was first raised in  mathematical physics
 several decades ago \cite{Faddeev1975}.
Faddeev proposed a model consisting of a three-component unit vector $\bfn=(n_1,n_2,n_3)$, $|\bfn|=1$:
\be
{\rm F} \ = \  \int d^3 {\bf r} \left\{\frac{1}{2}(\partial \bfn)^2 + 
c^2\left( \epsilon_{kij} \bfn \cdot \partial_i
\bfn \times \partial_j\bfn
\right)^2 \right\}
\la{fadmod}
\ee 
where  $ \epsilon_{kij}$ is the Levi-Civita symbol.
This model  supports topological defects in the form of closed or knotted loops (called knotted solitons) characterized by a nontrivial Hopf invariant  \cite{Faddeev1975}.
For such a defect in eq. (\ref{fadmod}), one can expect the energy coming from the first term (second order in derivatives integrated over the space) to scale as $[r]$ while the energy coming from  the last term (which is fourth order in derivatives) to  scale as $[1/r]$. 
For this reason it was argued that for this  soliton there is an energetically preferred length scale set  by the coefficient $c$
 \cite{Faddeev1975,knots,lin}.
Numerically the solutions for these defects were found only a decade ago by Faddeev and Niemi, confirming that 
the energy of a knotted soliton \cite{knots} has a global minimum at a certain length scale.
This was followed by a decade of highly nontrivial numerical and mathematical  studies 
 which uncovered a number of extremely interesting properties of these defects \cite{knots2} (for movies from numerical simulations by Hietarinta and Salo see ref. \cite{movies}).

The outstanding question is whether such topological defects  can be found in condensed matter systems.
In this case  several principal problems with realizability of these solitons 
can be immediately identified. First of all the nonmontonicity of the energy originates in a competition between second-order  and fourth-order derivative terms in an  energy functional. In quantum fluids, 
the Ginzburg-Landau or Gross-Pitaevskii energy functionals are effective models where a second order gradient term arises
from microscopic considerations via a derivative expansion. If one would try 
to stabilize knotted solitons by obtaining higher order terms  in a derivative expansion, the stabilization length would be such that 
the second and fourth order terms would be of the same order of magnitude as sixth-order and higher terms. Therefore the derivative expansion fails  and no Ginzburg-Landau description exists at this length scale. Thus
for a realization of the model like (\ref{fadmod}), the fourth-order derivative term should  have a {\it nonperturbative} origin.
That is, there should exist a regime where second- and fourth-order terms have
similar magnitude, at the same time being much larger than any higher-order terms.
 In \cite{PRL02} it was observed that the two-component Ginzburg-Landau (TCGL) model
 can be mapped onto a model containing the terms  (\ref{fadmod}). There, a fourth order term in derivatives  originates nonperturbatively
 as a contribution to magnetic field energy density.
However in contrast to eq. (\ref{fadmod}) the fourth-order term is coupled to another field.
The role of that complicated coupling to the additional massive vector field was not known. This question has also 
turned out to be very difficult to address numerically
because the studies of these solitons in the TCGL model are very computationally demanding.
The  numerical work, performed so far,  explored a limited range of parameters \cite{Juha} (see also the early work where however a dimensionality-reducing axially symmetric ansatz was used \cite{Ward}).
These first numerical works did not find  indications of the overlap of properties of the TCGL
model and the Faddeev model (\ref{fadmod}), even though 
 the results of the model (\ref{fadmod})  
were recovered by introducing a constraint 
which straightforwardly suppresses the additional massive vector field.
On the other hand these simulations did not rule out that there is a parameter range where knotted solitons are stable in the TCGL model and the question remained open. 
In this work I address this problem.

Let me briefly outline the mapping  \cite{PRL02} of  the TCGL model onto a model containing a version of  eq.  (\ref{fadmod}). In the simplest form, (used in \cite{PRL02}) the TCGL energy density is:
\be
F=\sum_{n=1,2}\frac{1}{2} |(\nabla+ie {\bf A})\Psi_n|^2 +V(|\Psi_n|)+ \f{1}{2} (\nabla \times{\bf A})^2 
\label{GL}
\ee
where $\Psi_n=|\Psi_n| e^{i \phi_n}$ are complex scalar fields which are coupled by the gauge field ${\bf A}$.
The symmetry breaking potential term $V$  can be quite general. Its role however is quite straightforward
to evaluate. Since in what follows we will focus on  the most interesting processes where the magnetic energy competes
against kinetic energy of superflow, the potential term will be used in the simplest $SU(2)$ form:
$V=v(|\Psi_1|^2+|\Psi_2|^2-[{\rm const}]^2)^2$ with a large coefficient $v$. In conclusion I briefly comment on the cases
of the effective potentials where the $SU(2)$ symmetry is broken.
The equation for supercurrent which follows from (\ref{GL})  is:
\be
{\bf J} =\f{1}{2}ie \sum_n [\Psi_n \nabla \Psi_n^*-\Psi_n^* \nabla \Psi_n] + e^2 \bfa |\Psi_n|^2
\label{super}
\ee
Lets introduce the following notations
\ba
&& \rho^2=|\Psi_1|^2+|\Psi_2|^2, \nonumber \\
&& \chi_n=|\chi_n|e^{i\phi_n}, \nonumber \\  
&& |\chi_n|=|\Psi_n|/\rho, \nonumber \\ 
&& |\chi_1|^2=\cos^2\left(\f{\theta}{2}\right); \ |\chi_2|^2=\sin^2\left(\f{\theta}{2}\right) \nonumber \\ 
&& \bfC={\bf J}/(e\rho^2), \nonumber \\
&& \bfj=i\sum_n[\chi_n \nabla \chi_n^*-\chi_n^* \nabla \chi_n]. \nonumber \\ 
\ea
Then we define the vector field 
\be
\bfn=(\cos(\phi_1-\phi_2)\sin\theta,\cos(\phi_1-\phi_2)\sin\theta,\cos\theta) 
\ee
for which the following identity holds
\cite{PRL02}:
\be
\f{1}{2} \rho^2 \left[|\nabla\chi_1|^2+|\nabla\chi_2|^2-\f{1}{2}\bfj^2\right] =\f{1}{8} \rho^2 (\nabla \bfn)^2.
\ee
The kinetic and magentic energy density terms of the model (\ref{GL}) then can be rewritten as \cite{PRL02}:
\ba
F&=&\f{\rho^2}{8}(\nabla\bfn)^2+\f{\rho^2}{2} \bfC^2 \nonumber \\
&+& \frac{1}{2e^2}\left[ \epsilon_{kij} \left(\partial_i C_j + \f{1}{4}\ndn\right)\right]^2 +V
\label{SV}
\ea
A knotted soliton in this model is defined as a texture of $\bfn$ characterized by a nontrivial Hopf invariant.
In the simplest case it is a ``vortex loop" which can be described as follows.
 Consider a cross section of this loop.
If at ${\bf r} \to \infty$ the value of $\bfn ({\bf r}) $ is $\bfn^0\equiv (n_1^0,n_2^0,n_3^0)$, then 
in the center of this cross section  we have $\bfn^c\equiv - \bfn^0 \equiv (-n_1^0,-n_2^0,-n_3^0)$.
In what follows I will call the point where  $\bfn^c\equiv - \bfn^0 \equiv (-n_1^0,-n_2^0,-n_3^0)$ the ``core",
even thought it does not have the same meaning as the core of an Abrikosov vortex.
Further, in this simplest case if we follow some path around the core where $\bfn \ne \bfn^0$ and  
$\bfn \ne \bfn^c$ a two component  vector $ {\bf l}$ (defined as a
 projection of $\bfn$ to a plane perpendicular to $\bfn^0$)  winds $N$  times along that path. Besides that 
 $ {\bf l}$ winds $M$ times along any closed paths in toroidal direction, i.e. along the core (that means that this closed 
 vortex is ``prepared" by twisting a skyrmion-like fluxtube $M$ times before gluing its ends).
Because of these windings, for any value of ${\bfn}$ one can identify one or several 
closed helices in the 
physical space where $\bfn=\tilde{\bfn}$ (which is called a preimage of $\tilde{\bfn}$).
The winding $N$ specifies the number of these helices 
while the winding $M$ specifies how many steps these helices have. In general case the vortex  can have
a form of a closed knot. Then the  preimages of $\tilde{\bfn}$  are closed knotted helices (see Fig. 1).
\begin{widetext}
\begin{center}
\begin{figure}[h]
  \includegraphics[width=\columnwidth]
  {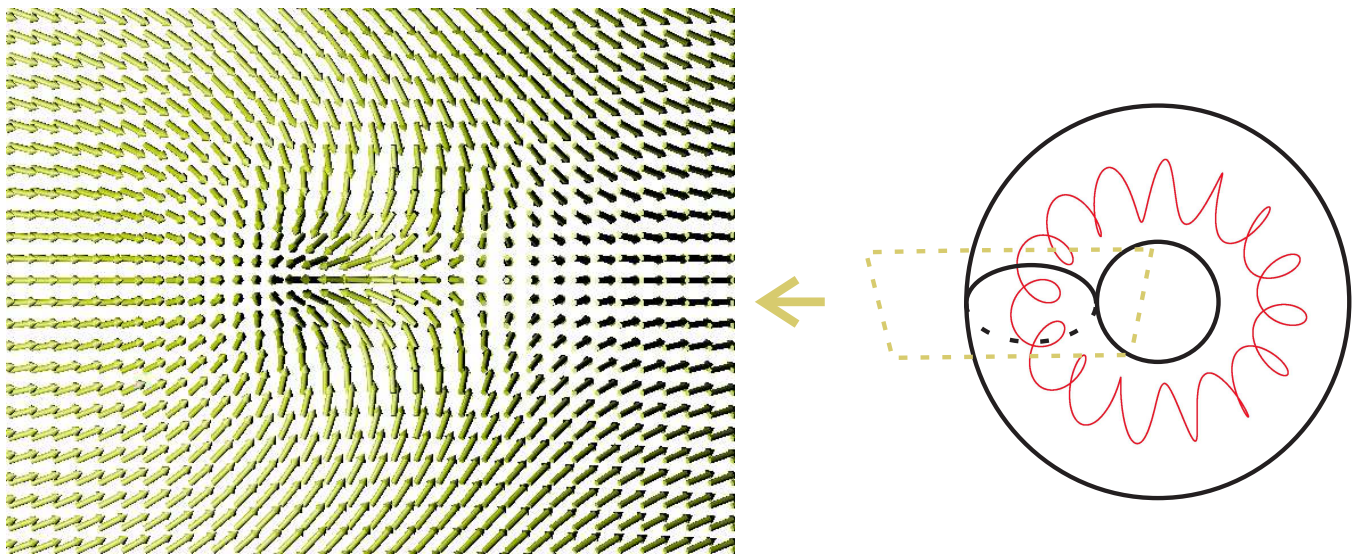}
  \caption{\label{1} (Color online)  A schematic picture of a toroidal knotted soliton (shown on the right). Any cross section of the 
flux tube (rectangular area, schematically shown in yellow) produces a skyrmionic texture of the three-component vector $\bfn$.
 The texture of $\bfn$, close to the vortex center in the cross section, is shown on the left. 
The fluxtube is twisted: the red helix schematically shows the preimage of the north pole: i.e. where in physical space $\bfn$ assumes 
the position corresponding to the ``north pole" on $S^2$ (i.e.$\bfn=(0,0,1)$ ) }
\end{figure}
\end{center}
 \end{widetext}
Consider a toroidal knot with diameter much larger than the penetration length  $\lambda\equiv {1}/{e\rho}$.
There is quantized magnetic flux carried by this vortex. This follows from the following argument.
Consider a vortex loops of size  $L\gg \lambda$. Consider a path $\sigma$ 
which is much smaller than $L$ such that it goes once around the core
at a distance much larger than $\lambda$ . 
Since by definition  such a path is outside the texture where $\bfn$ has windings or varies appreciably,
we have everywhere along  that path $\bfn\approx\bfn^0$ (on the left side of the Fig. 1 it would
correspond to a contour which is far enough from the core so that  along that contour, the vector $\bfn$ does not deviate from the 
value on equator in the order parameter space $S^2$ and pointing to the right). From there is follows that along this path
$\oint_\sigma \nabla\beta=0$ and $\oint_\sigma \nabla(\phi_1-\phi_2)=0$.
Further if a cross section of the flux tube is characterized by the winding number $N$, it should 
contain $N$ preimages of the north pole of $S^2$ (defined as a point in the cross section where 
$\theta=0$ and thus $|\Psi_1|=0$) as well as $N$ preimages of the south pole of $S^2$ 
(defined as a point in the cross section where  $\theta=\pi$ and thus $|\Psi_2|=0$).
{\it The only proper mapping from Ginzburg-Landau variables $\Psi_i, {\bf A}$ to the variables $\rho,\bfn,\bfC$
should be supplemented by imposing the condition that singlevaluedness of $\Psi_n$ is preserved.}
This condition is: along any path $\mu$ in the physical space which encircles $N_1$
preimages of the north pole of $S^2$ and $N_2$ preimages of the south pole of $S^2$ the conditions
should hold $\oint_\mu \nabla \phi_1=2\pi N_1$ and $\oint_\mu \nabla \phi_2=2\pi N_2$. Note that 
to be consistent with a Hopf map each of the phases should have a single $2\pi$ winding per zero $N_n$
of the order parameter $|\Psi_n|$, for all boundary conditions, except the case where $\bfn$ assumes positions corresponding  to the north or south pole on $S^2$ at infinity and in the core.
Thus using the additional singlevaluedness conditions we have for the path $\sigma$:
\be
\oint_\sigma \nabla (\phi_1-\phi_2)=0; \ \ \oint_\sigma \nabla (\phi_1+\phi_2)= 4\pi N
\ee
This determines the magnetic flux enclosed by the fluxtube, as follows from the equations of motion (\ref{super}):
\ba
&&\Phi= \oint_\sigma ds {\bf A} = \nonumber \\
&&\f{1}{e^2 (|\Psi_1|^2+ |\Psi_2|^2)}\oint_\sigma ds \left[ {\bf J}
-\f{ie}{2} \sum_n [\Psi_n \nabla \Psi_n^*-\Psi_n^* \nabla \Psi_n] \right] \nonumber \\
&& \approx \f{1}{e^2 (|\Psi_1|^2+ |\Psi_2|^2)}\oint_\sigma ds \left[ 
-\f{ie}{2} \sum_n [\Psi_n \nabla \Psi_n^*-\Psi_n^* \nabla \Psi_n] \right]\nonumber \\
&&=-\Phi_0 N.
\ea
where $\Phi_0=\f{2\pi}{e}$  is the magnetic flux quantum. 
%

 Note that the flux quantization of a topologically nontrivial texture of $\bfn$
appears only if the model (\ref{SV}) is supplemented with the additional 
conditions of singlevaluedness of the original phases $\phi_n$. Without this 
extra conditions the model (\ref{SV}) allows topologically nontrivial textures of
$\bfn$ which do not carry magnetic flux.  It is thus an important 
circumstance that the 
model (\ref{SV}) has configurations which do not correspond to physical configurations in 
(\ref{GL}), { therefore the proper mapping from (\ref{GL}) to (\ref{SV})  should involve the constrain
associated with the implementation of the singlevaluedness conditions on $\Psi_n$.}
Note also the texture carries one flux quantum per couple of spatially separated zeros of $\Psi_1$ and $\Psi_2$,
therefore a knotted soliton can be interpreted as a special bound state of twisted fractional flux vortices \cite{frac}.

The described above topological objects  are always well defined when $\rho \ne 0$ at any point in that texture. The
more complicated question is how the energy of these defects scales as a function of their sizes. I.e. if these defects 
 can minimize their energy by shrinking or if they will be protected from shrinking by an energy barrier.
Lets return to the eq. (\ref{SV}). The last term is the most interesting here. It represents the magnetic field energy density
which  has the contributions
from the massive vector field $\bfC$ and also a Faddeev term $[\epsilon_{kij}\ndn]^2$.  It suggests that, 
if there are conditions where the contribution from $\nabla \times \bfC$ is negligible, then
 the magnetic field energy density in the TCGL model would scale as a Faddeev  term.
However, at least in the limit $e \to 0$, when ${\bf B}\to 0$ 
one finds that $\e \partial_i C_j =- \f{1}{4} \e\ndn$  and the model is reduced to a Gross-Pitaevskii equation for two decoupled fields
without a self-generated Faddeev term. Moreover, similar  behavior 
has  been seen in the recent  numerical studies \cite{Juha} of the eq. (\ref{GL}) for a range of parameters. 
This raises the question whether the electrodynamics of two-component superconductor actually possess a self-generated Faddeev  term or if it is compensated by the field $\bfC$ making scaling properties of vortex loops monotonic.
Below I develop analytic treatment for the role of the field $\bfC$ to answer this question.

Let me first observe that if there exists a regime  where the system tends to the model (\ref{fadmod})
(i.e. where the field $\bfC$ plays a role of small correction),
 then a {\it self-consistent} perturbative scheme for treating the field $\bfC$ is possible.
The vector $\bfn$ indeed depends  only on gauge invariant quantities such as the  phase difference
between the condensates and the relative densities. Therefore one can always define 
a texture of $\bfn$, and the accompanying configuration of the field  $\bfC$ will be determined by a configuration of the vector potential 
corresponding to the energy minimum of the functional (\ref{GL}) for a given texture of $\bfn$. 
Consider now a texture of $\bfn$ corresponding to a knotted soliton. 
Further consider shrinking of that texture.
In this process the contribution from $\e\ndn$ in the energy functional will grow. 
The field $\bfC$ should assume then an optimal 
configuration from the point of view of the
energetic balance between (i) the best possible compensation of the $\e\ndn$ contribution
 in the last term in (\ref{SV}),
and (ii) the accompanying energy penalty in the second term in (\ref{SV}). 
Observe now that if  there exists a parameter regime and a characteristic size of a knotted soliton 
 where the self-generated 
 magnetic field  consists primarily of the Faddeev  term contribution: 
 $ {B_k} \approx \f{1}{4e}\e\ndn$, then
a contribution from  the field $\bfC$ can be estimated perturbatively.
Since it is not {\it a priory} known if such a regime exists, the perturbative scheme 
should be {\it self-consistent}. That is, the breakdown of the self-consistency criterion should signal the violation of the assumption.
Such a self-consistent perturbative estimate of the contribution from the field $\bfC$ 
can be made by using the condition that the Maxwell equation ${\bf J} = \nabla \times {\bf B}$  should be satisfied.
First observe that from ${\bf J} = \nabla \times {\bf B}$ it  follows that:
\be
\bfC = \lambda^2 \left\{\nabla \times \nabla \times \bfC +\f{1}{4} \nabla \times \e \ndn\right\},
\label{fc}
\ee
and
\ba
&&\nabla \times\bfC =\lambda^2 \Biggl\{\nabla \times \nabla \times \nabla \times \bfC +
 \nonumber \\
&&\f{1}{4}\nabla \times \nabla \times [\e\ndn] \Biggr\} \nonumber
\ea
where $\lambda\equiv {1}/{e\rho}$. 
By substituting this into  the expression for ${\bf B}^2$ and repeating the procedure 
iteratively,  an expansion in powers of $\lambda$ can be generated:
\ba
\f{{\bf B}^2}{2} &\approx & \f{1}{32e^2}
\Bigl\{
 \e\ndn  + \nonumber \\&&{\lambda^2}\nabla \times\nabla \times[\e\ndn]
+ ...
\Bigr\}^2
\ea
Applying a similar expansion for the term $(\rho^2/2) \bfC^2$ with the help of eq. (\ref{fc}) 
it follows that when $\lambda$ is much smaller than the characteristic size $L$  of the texture of $\bfn$
the dominant terms in the model are:
\be
F^{[L\gg  \lambda]}\approx\f{\rho^2}{8}(\nabla\bfn)^2 + \frac{1}{32e^2}\left(\e\ndn\right)^2
\label{largescale}
\ee
On the other hand for  $L \approx \lambda$ the above self-consistent  exclusion of the field $\bfC$ fails, and in the limit 
$L/\lambda  \to 0$ one approaches the $e=0$ scaling where $\nabla \times \bfC\approx - \f{1}{4} \ndn$,
as can be seen from the equation
\ba
&&\f{\bfC}{\lambda^2} =\left\{\nabla \times \nabla \times \bfC +\f{1}{4} \nabla \times \e \ndn\right\} \nonumber \\
&&\to 0 \ (\lambda \to \infty)
\ea
From here a conclusion follows that a texture characterized by a nontrivial Hopf in the TCGL model
with size much smaller than $\lambda$ can minimize its energy by shrinking.
It means that for a given Hopf charge a knot soliton in (\ref{GL}) can represent at most a local
energy minimum, while the knotted solitons in the original Faddeev model always correspond to a global
minimum (see also remark \cite{remark3}). 

However a knotted soliton texture which is  much larger than  $\lambda$ should have energy scaling 
similar to that in the model (\ref{fadmod}), i.e. receiving a contribution $~(\e\ndn)^2$ in the  energy density.
Considering the case of the lowest Hopf charge $Q$, the energy functional (\ref{largescale}) has stable solitons 
with characteristic size given by the ratio between the second- and fourth-order derivatives terms
which is $L_0=1/(2e\rho)=\frac{1}{2}\lambda$, note however that the self-consistent procedure of excluding the 
field $\bfC$, which was used to obtain (\ref{largescale}),  breaks down  at the scales of order of $\lambda$,
(where the energy of a knot texture is affected significantly by the coupling to  $\bfC$).
The lack of small parameters  in this regime makes it  difficult to estimate energy scaling. 
However it is  possible that there is a {finite} potential
barrier at this length scale which can prevent a texture from a shrinkage. 
The observation of the 
instability of knotted solitons in the TCGL model in the recent numerical studies \cite{Juha}
may originate from the simulations being in the parameter range of the monotonic scaling regime.
This will be the case if for example the initial texture is too small compared to $\lambda$, or 
it can also result from $\lambda$ being too small compared to the numerical grid spacing
making the stabilization length being too small to resolve on a numerical grid etc.
Indeed finding a finite potential barrier in numerical simulations is a much harder task than the identification of the 
infinite energy barrier in the original Faddeev model  
 (where a texture scales to the global minimum no matter what are the initial conditions). 
Finding a small potential barrier with procedures like in 
 \cite{Juha} requires fine-tuning of the initial texture to be close enough to the one corresponding to
the local minimum.

Importantly,  the above considerations are
restricted to solitons with lowest Hopf charges $Q$. However, one of the most
 remarkable properties
 of the model (\ref{fadmod}) is the
existence of a Vakulenko-Kapitanskii bound,
which states that energy of a soliton depends on the  Hopf charge $Q$
 as $E\geq const \cdot c|Q|^{3/4}$ \cite{Vakulenko}. Because of the $3/4$ power,  solitons with high Hopf charge are stable against decay into several solitons
 with lower Hopf charge. Indeed high-$Q$ solitons have size which depends not only on the length scale 
 given by the ratio of the coefficients in front of the second- and fourth-order terms in (\ref{SV}) but also on Hopf charge.
The textures should in general be larger for larger $Q$, (though there is no simple scaling because at high-Q
solitons develop very complicated forms \cite{knots2}). It indicates 
a possibility  that in case of a high Hopf charge, knotted solitions in the TCGL model 
may have a local energy minimum at the length scales larger than $\lambda$, and thus
be reasonably well described by the effective model (\ref{largescale}). 

The complicated nature of the energy scaling of the knotted soliton texture in the simplest GL model (\ref{GL})
is connected with the fact that it has only one length scale $\lambda$.
This scale sets the ratio between second-order and forth-order derivative terms in $\bfn$
and at the same time it sets the inverse mass for the vector field $\bfC$. In general physical systems the TCGL model includes 
other terms consistent with the symmetry. One very generic term which is second order in derivatives
is the intercomponent current-current interaction (the Andreev-Bashkin terms) \cite{AB}. 
TCGL with these terms is:
\ba
F&=&\sum_{n=1,2}\frac{1}{2} |(\nabla+ie {\bf A})\Psi_n|^2+ V(|\Psi_n|)+ \f{1}{2} (\nabla \times{\bf A})^2 \nonumber \\
&+& \alpha\Psi_1^*\Psi_2^*(\nabla+ie {\bf A})\Psi_1\cdot(\nabla+ie {\bf A})\Psi_2  \nonumber \\
&+& \alpha\Psi_1\Psi_2(\nabla-ie {\bf A})\Psi_1^*\cdot(\nabla-ie {\bf A})\Psi_2^* \nonumber \\
&-& \alpha\Psi_1\Psi_2^*(\nabla-ie {\bf A})\Psi_1^*\cdot(\nabla+ie {\bf A})\Psi_2  \nonumber \\
&-& \alpha\Psi_1^*\Psi_2(\nabla+ie {\bf A})\Psi_1\cdot(\nabla-ie {\bf A})\Psi_2^* 
\label{GL2}
\ea
In physical systems the coefficient $\alpha$ can vary in a wide range and be either negative or positive.
Let us consider electrodynamics and  knotted solitons stability in the model (\ref{GL2}).
A new separation of variables can be introduced by defining the following fields:
\ba
\cho&\equiv& \sin\f{\tilde{\theta}}{2} e^{i\phi_1}, \  \cht\equiv \cos\f{\tilde{\theta}}{2}e^{i\phi_2}, \nonumber \\ 
\bfe &\equiv&  (\cho, \cht) \vec{\sigma}
  \begin{pmatrix} \cho^*\\ \cht^* \end{pmatrix} \nonumber \\
\cho &=&\sqrt{\frac{|\Psi_1|^2 (1-4\alpha|\Psi_2|^2)}{ (1-4\alpha|\Psi_2|^2)|\Psi_1|^2 +(1-4\alpha|\Psi_1|^2)|\Psi_2|^2}}e^{i\phi_1}, \nonumber \\
\cht &=&\sqrt{\frac{|\Psi_2|^2 (1-4\alpha|\Psi_1|^2)}{ (1-4\alpha|\Psi_2|^2)|\Psi_1|^2 +(1-4\alpha|\Psi_1|^2)|\Psi_2|^2}}e^{i\phi_2}, \nonumber \\
{\bfCt}&=&
\Biggl[\f{i}{2} \Bigl\{\cho \nabla \cho^*-\cho^*\nabla \cho +\cht  \nabla \cht^*-\cht^*\nabla \cht \Bigr\}  
+e \bfa \Biggr]; \nonumber \\
\rhot^2 &\equiv& (1-4\alpha|\Psi_2|^2)|\Psi_1|^2 +(1-4\alpha|\Psi_1|^2)|\Psi_2|^2; 
\label{sc}
\ea
Lets, as in the previous example, consider the regime where one can neglect density fluctuations except the relative density fluctuations
described by $\tilde{\theta}$  (that is,  we will be working with the  $O(3)$ field $\bfe$ coupled 
to a massive vector field $\bfCt$).Then the model can be rewritten as:
\ba
&&F\approx \f{\rhot^2}{8}(\nabla\bfe)^2+\f{\rhot^2}{2} \bfCt^2 +
\nonumber \\
&& \frac{1}{2e^2}\left[ \nabla\times \bfCt
 + \f{1}{4}\e \ede\right]^2+
\nonumber \\
&&\alpha \rhot^4 \frac{|\cht \nabla \cho^* + \cho^* \nabla \cht |^2 + |\cht \nabla \cho - \cho \nabla \cht|^2 }{(1-4\alpha|\Psi_1|^2)(1-4\alpha|\Psi_2|^2)} 
\label{SV2}
\ea
From here it follows that the model (\ref{GL2}) can be represented as the model (\ref{SV}) with a renormalized 
 characteristic length scale $\lt \equiv{1}/{(e \rhot)}$ and the additional term (the last term in (\ref{SV2})).
This term plays a crucial role, namely it breaks the single-parameter character of the TCGL model (\ref{GL}).
This  follows from the following  identity:
\ba
&&\f{\rhot^2}{8}(\nabla\bfe)^2+ \nonumber \\
&& \alpha \rhot^4 \frac{|\cht \nabla \cho^* + \cho^* \nabla \cht |^2 + |\cht \nabla \cho - \cho \nabla \cht|^2 }{(1-4\alpha|\Psi_1|^2)(1-4\alpha|\Psi_2|^2)} =
\nonumber \\
&&\f{\rhot^2}{8}\Bigl[ (\nabla \tilde{\theta})^2 + \sin^2 \tilde{\theta}(\nabla (\phi_1-\phi_2))^2 \Bigr] \nonumber \\
&&+ \frac{\alpha \rhot^4}{2(1-4\alpha|\Psi_1|^2)(1-4\alpha|\Psi_2|^2)} \times \nonumber \\ 
&&\Bigl[\f{1+ \cos^2\tilde{\theta}}{2} (\nabla \tilde{\theta})^2 + \sin^2 \tilde{\theta}(\nabla (\phi_1-\phi_2))^2 \Bigr]
\ea
For $\alpha<0$ the second term diminishes the energy coming from the term $\f{\rhot^2}{8}(\nabla\bfe)^2$.
For this reason the second-order derivative terms are balanced by  the fourth order term $\sim  (\e\ede)^2$
at a larger texture size (relative to the scale $\lt$ associated with the field $\bfCt$).
Thus the model  (\ref{GL2}) has a tunable disparity of the characteristic lengths for $\bfe$ and $\bfCt$. 
Therefore for a large enough negative $\alpha$ a knotted 
soliton in the model (\ref{GL2}) should be stabilized at the length scale much larger than $\lt$ and thus approach the properties of the knotted solitons
in the model (\ref{fadmod}).

In conclusion,  normally the essence of electrodynamics of a superconductor is understood as the fact that the phase field
and vector potential combine to produce a massive vector field which then describes the electrodynamics of the system.
In this work the electrodynamics of a two-component superconductor is considered 
and it is shown that a two-component superconductor not only has  distinct electrodynamics manifested in the generation of a Faddeev term along with a massive vector field, but also that there are regimes where the electrodynamics is dominated by this term.
There is a crossover to smaller  length scales
where effectively the electrodynamics does not feature the contribution in the form of the Faddeev term. Therefore
the knotted solitons  in the  TCGL model may be a local minimum in the energy for a given Hopf charge (in contrast to 
a global minimum in the case of the model (\ref{fadmod})). 
In the second part of the paper I showed that a more generic TCGL model
 with physically relevant mixed gradient terms  possesses two characteristic length scales.
This makes the potential barrier for knotted solitons tunable.
Such vortex loops, which have a potential barrier against shrinkage, should lead to entirely different quench reaction,
superfluid turbulence and physics of thermal fluctuations. 
The results may be  relevant for a  variety of physical systems where two-component Ginzburg-Landau model is realized
ranging from electronic multicomponent  superconductors to the projected mixtures of protonic and electronic 
condensates in liquid metallic hydrogen \cite{Nature}.
If in the models like (\ref{GL}), the knotted solitons being a local minimum of the energy functional, 
have energy proportional to $Q^{3/4}$ like their kin in the Faddeev model, then these defects, if e.g. induced by fluctuations will have tendency to pileup. Such a behavior may be relevant for  understanding the
recent observations in numerical simulations of a discontinuous phase transitions 
in $SU(2)$-superconductor \cite{DQC1}.
In connection with applicability to physical systems, it should also be noted that in principle it is 
not necessary to have the exact $SU(2)$ symmetry for the realization of the physics discussed above.  Because a knotted soliton 
is a closed loop, it does not produce any phase windings at infinity, therefore an 
effective potential  which breaks $SU(2)$ symmetry to $U(1)\times U(1)$ (or even softly breaks it to $U(1)$) introduces
only a finite energy penalty for a knotted soliton and does not necessarily destroys it. 
 
 I thank Juha Jaykka for the plot of the cross section of knot soliton.

\end{document}